\documentclass[letter,column]{jpsj3}
\usepackage{amsmath}
\usepackage{amssymb}
\usepackage{graphicx}
\usepackage[usenames,dvipsnames]{color}
\usepackage{txfonts}

\title{Open Quantum Dynamics Theory for Non-Equilibrium Work: Hierarchical Equations of Motion Approach}

\author{Souichi Sakamoto and Yoshitaka Tanimura$^*$}
\inst{Department of Chemistry, Graduate School of Science, Kyoto University, Kyoto 606-8502, Japan} 

\abst{A system--bath (SB) model
is considered to examine the Jarzynski equality in the fully quantum regime.
In our previous paper [J. Chem. Phys. \textbf{153} (2020) 234107], we carried out ``exact'' numerical experiments using hierarchical equations of motion (HEOM) in which we demonstrated that the SB model describes behavior that is consistent with the first and second laws of thermodynamics and that the dynamics of the total system are time irreversible. 
 The distinctive quantity in the Jarzynski equality is the ``work characteristic function (WCF)'', $\langle \exp[-\beta W] \rangle$, where $W$ is the work performed on the system and $\beta$ is the inverse temperature.
In the present investigation, we consider the definitions based on the partition function (PF) and on the path, and numerically evaluate the WCF 
using the HEOM to determine a method for extending the Jarzynski equality to the fully quantum regime. We show that using the PF-based definition of the WCF, we obtain a
 result that is entirely inconsistent with the Jarzynski equality, while if we use the path-based definition, we obtain a result that approximates the Jarzynski equality, but may not be consistent with it.}

\begin{document}
\maketitle

In thermodynamics, work, $W$, and heat, $\Delta Q$, are thermodynamic process quantities, while the internal energy, $\Delta U$, is an extensive quantity that cannot be measured directly. For a classical system beginning in an equilibrium state, it has been found that the work done under an arbitrary mechanical operation is related to the equilibrium free energy in accordance with the Jarzynski equality\cite{JarzynskiPRL97,Jarzynski04,JarzynskiAnnu11,CrooksPRE99,SeifertPRL05,EvansPRL08,MaiPRE07,SeifertJSM07,SaitoPRB08,Liphardt02,BlickePRL06,DouarcheEuro05,RitortJCP09,CollinNature05}: $-\ln \left( \langle \exp[-\beta W(\tau)) \rangle \right)/\beta =\Delta F_A(\tau)$. Here, $\beta \equiv 1/k_{\mathrm{B}}T$ is the inverse temperature with the Boltzmann constant $k_\mathrm{B}$, $\Delta F_A(\tau)$ is the change in the free energy of the system, $W(\tau)$ is the non-equilibrium work, and $\langle \ldots \rangle$ is the ensemble average over all phase space trajectories under the time-dependent external perturbation from time $t=0$ to $\tau$.  
Although investigating  this equality  in the classical regime is straightforward in both theoretical and experimental contexts, doing so in the quantum regime remains challenging,\cite{Hanggi2020,Mukamel,Campisi11,Campisi09,CrooksStat08,YukawaJPSJ00,Tasaki,Kurchan,Alonso} because 
the dynamics of a small quantum system itself are reversible in time and therefore the system cannot reach thermal equilibrium on its own without a system--bath interaction: We cannot assume a canonical distribution as the equilibrium state for either the system of the heat-bath, due to the presence of the system-bath interaction.  In the present paper, we numerically evaluate the work characteristic function (WCF), $\langle \exp[-\beta W(\tau)] \rangle$, and $\Delta F_A(\tau)$ with the goal of extending the Jarzynski equality to the fully quantum regime. 

A commonly employed model for this kind of investigation is described by a system-bath (SB) Hamiltonian, in which a small quantum system $A$ is coupled to a bath $B$ modeled by an infinite number of harmonic oscillators. We found that the behavior described by this model is consistent with the first and second laws of thermodynamics and provides an ideal platform to examine various fundamental propositions of thermodynamics in the fully quantum regime.\cite{KATO2016,Sakamoto2020JCP}
In this model, the Hamiltonian of the total system is given by
\begin{align}
    { {\hat H}}(t) = {{\hat H}}_A(t) + {{\hat H}}_I + {{\hat H}}_B,
    \label{eq:Hamiltonian}
\end{align}
where $ {{\hat H}}_A(t)$, $ {{\hat H}}_I$ and $ { {\hat H}}_B$ are the Hamiltonian of the system, the interaction and the bath, respectively. The system Hamiltonian is given by ${{\hat H}}_A(t) = {{\hat H}}_A^0 + { {\hat H}}_E(t)$, with ${{\hat H}}_A^0 = \frac{1}{2} \hbar \omega_0 (|e \rangle \langle e| - |g \rangle \langle g|)$ and ${{\hat H}}_E(t)=0$ for $t \le 0$, where $|e\rangle$ and $|g\rangle$ are the excited and ground states of the system, and ${{\hat H}}_E(t)$ is the interaction Hamiltonian with an external field. 
The bath Hamiltonian ${{\hat H}}_B$ is expressed as
\begin{align}
    {\hat H}_B = \sum_j \left[ \frac{\hat p_j^2}{2 m_j} + \frac{1}{2} m_j \omega_j^2 {\hat x_j}^2 \right],
    \label{eq:bathHamiltonian}
\end{align}
where $\hat p_j$, $\hat x_j$, $m_j$ and $\omega_j$ are the momentum, position, mass and frequency of the $j$th bath oscillator, respectively. The SB interaction ${ {\hat H}}_I$ is given by ${ {\hat H}}_I ={\hat V} \sum_j g_j {\hat x}_j$, where $\hat V$ is the system part of the interaction,
and $g_j$ is the coupling constant between the system and the $j$th bath oscillator. 
The effect of the bath is characterized by the noise correlation function, $C(t) \equiv \langle{\hat X}(t) {\hat X}(0) \rangle_B$, where ${\hat X} \equiv \sum_j g_j {\hat x}_j$, and $\langle \ldots \rangle_B$ represents the average taken with respect to the canonical density operator of the bath. The noise correlation function is expressed as
\begin{align}
    C(t) = \hbar \int_0^{\infty} \frac{d \omega}{\pi} J(\omega) \left[ \coth \left( \frac{1}{2}\beta \hbar \omega \right) \cos(\omega t) - i \sin(\omega t) \right],
    \label{eq:correlation}
\end{align}
where $J(\omega) \equiv \sum_{j} \left({\pi g_{j}^2}/{2m_{j} \omega_{j} }\right) \delta(\omega-\omega_{j})$ is the spectral density and $\beta$ is the inverse temperature of the bath. 

When we apply the SB model to problems of thermodynamics, because the main system is microscopic and because the quantum coherence between the system and bath characterizes the quantum nature of the system dynamics, the role of the SB interaction has to be examined carefully. For example, although the factorized thermal equilibrium state, $\hat \rho_{tot}^{eq}= \hat  \rho_A^{eq} \otimes \hat \rho_B^{eq}$, where $\hat  \rho_A^{eq}$ is the equilibrium state of the system without the SB interaction, is often employed as an initial state when investigating open quantum dynamics, in actual situations, the system and bath are quantum mechanically entangled (a phenomenon referred to as ``bath entanglement'').\cite{YTperspective,TanimuraJPSJ06} 

In Refs. \citeonline{KATO2016} and \citeonline{Sakamoto2020JCP}, we presented a scheme for calculating thermodynamic variables in the SB model on the basis of simulations including an external perturbation using the hierarchical equations of motion (HEOM).\cite{YTperspective,TanimuraJPSJ06,Tanimura89A,TanimuraPRA90,IshizakiJPSJ05,YTJCP14,YTJCP15}   The key quantity in this investigation is the change of the ``quasi-static Helmholtz energy'' at time $\tau$, which is defined as\cite{Sakamoto2020JCP}
\begin{align}
    \Delta F_A(\tau) \equiv \int_0^{\tau} {\operatorname{tr}_A} \left\{ {\hat \rho}_A^{qeq} (t) \frac{\partial}{\partial t} { {\hat H}_A}(t) \right\} dt,
    \label{eq:Fchange}
\end{align}
where  ${\hat \rho}_A^{qeq} (t) $ is the ``quasi-static'' reduced density operator. Here, note that, as we demonstrated numerically, when ${\hat H}_A (t)$ changes much more slowly than the relaxation time of the system, $\hat \rho_A (t)$ can be evaluated within the HEOM approach as the quasi-thermal equilibrium state of the system, ${\hat \rho}_{A}^{qeq} (\tau) \approx {\operatorname{tr}_{B}} \{ e^{-\beta({ {\hat H}_A}(\tau) + {\hat H}_I + {{\hat H}_B})} \}/ Z_{tot} (\tau)$, where  $Z_{tot} (\tau) \equiv {\operatorname{tr}_{A+B}} \{ e^{-\beta({ {\hat H}_A}(\tau) +  {\hat H}_I + {{\hat H}_B})} \}$ at time $\tau=t$.  Then the change of the ``quasi-static Boltzmann entropy'', $\Delta S_A(\tau)$, is given by
\begin{align}
    \Delta S_A(\tau) = k_B \beta^2 \frac{\partial}{\partial \beta} \Delta F_A(\tau).
    \label{eq:Bchange}
\end{align}
Although the increase of the internal energy and the Boltzmann entropy of the system arise not only from the change in the system Hamiltonian itself but also from the change in the system part of the SB interaction, the HEOM allows us to evaluate both variables accurately.  Using the HEOM, we can also evaluate the change of the bath part of the SB interaction energy and bath energy, while these energies themselves are treated as infinitely large, because the bath is regarded as possessing an infinitely large heat capacity. With this treatment, we found that the SB model describes behavior that is consistent with the first law of thermodynamics. Explicitly, we found that the relation, $\langle W(\tau) \rangle =\Delta U_A(\tau) -\Delta Q(\tau)$, is satisfied, where  $\Delta U_A(\tau)= \Delta F_A(\tau)+ T\Delta S_A(\tau)$ is the internal energy of the system and $\Delta Q(\tau)$ is the heat released from the bath.\cite{Sakamoto2020JCP} Moreover, we have numerically confirmed that  the total entropy production is alway positive. With these results strongly supporting the validity of our approach, in this paper, we evaluate $\langle \exp[-\beta W(\tau)] \rangle$ using the HEOM formalism with the goal of extending the Jarzynski equality to the fully quantum regime characterized by a non-Markovian and non-perturbative SB interaction. 

In what follows, we examine two definitions of the WCF:  (i) a definition based on the partition function (PF) (the PF-WCF)\cite{Talkner16} and (ii) a definition based on trajectory (path) (the path-WCF).
For an isolated quantum system, the PF-WCF has been defined as\cite{HanggiW}
\begin{align}
\langle \exp[-\beta W(\tau)] \rangle_{PF} &\equiv tr\left \{  {\rm e} ^{-\beta {\hat H}^{(H)} (\tau) }  {\rm e} ^{\beta {\hat H} (0) } {\hat \rho}_{tot}(0) \right\} \nonumber\\
 &= \frac{Z_{tot}(\tau)}{Z_{tot}(0)},
\label{eq:Hanggi}
\end{align}
where $ {\hat \rho}_{tot}(0) \equiv {\hat \rho}_{tot}^{eq}$, and ${\hat H}^{(H)}(\tau)$ is the Heisenberg operator of $ {\hat H}(\tau)$.
We rewrite this expression in terms of the time-reversal Liouville operator as $\langle \exp[-\beta W(\tau)] \rangle_{PF}={\rm tr} \left\{{\hat P} _{PF}(\tau) \right\}$, where
${\hat P}_{PF} (\tau) = \exp_+ \left [   -i \int_{\tau}^0 dt {\hat H}_{tot}^{\times}(t)/ {\hbar}  \right]  {\hat Z}_{tot}(\tau)$, with $ {\hat Z}_{tot}(\tau) \equiv\exp\left[ -\beta {\hat H} _{tot}(\tau) \right]$, and we have introduced the time-ordered exponential  ${\rm exp}_{\pm}$.  Here and hereafter we use the hyperoperator notation ${\mathcal {\hat O}}^{\times} {\hat f} \equiv {\mathcal {\hat O}}{\hat f}-{\hat f}{\mathcal {\hat O}}$ and ${\mathcal {\hat O}}^{\circ} {\hat f} \equiv {\mathcal {\hat O}}{\hat f}+{\hat f}{\mathcal {\hat O}}$ for any operator ${\mathcal {\hat O}}$ and operand operator ${\hat f}$. Unfortunately, because the heat bath possesses infinitely many degrees of freedom,  we cannot evaluate ${\hat P}(\tau)$.  Instead, because $\langle \exp[-\beta W(\tau)] \rangle_{PF}=\left(Z_{A}(\tau)Z_{B}(\tau)\right)/\left(Z_{A}(0) Z_{B}(0)\right)$, where $Z_{A}(\tau)$ and $Z_{B}(\tau)$ are the system and bath parts of the partition functions, we can evaluate it indirectly using Eq.\eqref{eq:Fchange} as $\langle \exp[-\beta W(\tau)] \rangle_{PF} \approx \exp \left[ -\beta(
\Delta U_A(\tau) -\Delta Q(\tau))\right]$ with  $\Delta U_A(\tau)= \Delta F_A(\tau)+ T\Delta S_A(\tau)$, which, off course, is the first law of thermodynamics.\cite{Sakamoto2020JCP}

Alternatively, we can use the path-WCF on the basis of the non-equilibrium trajectories (paths) as  $ \langle \exp[-\beta W(\tau)] \rangle_{path}={\rm tr} \left\{ {\hat K}(\tau) \right\}$, where 
 ${\hat K}(\tau) \equiv {\hat U}(\tau, 0) {\hat P}_{path}(\tau) {\hat U}^{\dagger}(\tau, 0)$, with 
\begin{align}
  {\hat P}_{path}(\tau) = {\rm e}_+ ^{-\frac{\beta}{2} \int_0^{\tau} dt {\dot {\hat H}_A^{(H)} }(t) }{\hat \rho}_{tot}(0) {\rm e}_- ^{ -\frac{\beta}{2} \int_0^{\tau} dt {\dot {\hat H}_A^{(H)}} (t) },
  \label{eq:P} 
\end{align}
and ${\hat U}(\tau, 0) \equiv \exp_+ \left[ -(i/\hbar)\int_{0}^{\tau} dt {\hat H} (t) \right]$.
Here, instead of ${\hat P}(\tau) =\exp_+ \left[ -\beta \int_0^{\tau} dt {\dot {\hat H}_A^{(H)}} (t) \right] {\hat \rho}_{tot}(0)$,\cite{HanggiW}   we use the symmetric form in Eq. \eqref{eq:P},  because otherwise $ {\hat P}(\tau)$ is not Hermitian and, as a consequence, the WCF may not be real valued.  The equation of motion for $ {\hat K}(\tau) $ is given by
\begin{align}
  \frac{\partial}{\partial \tau} {\hat K}(\tau) = -\left[ \frac{i}{\hbar}{\hat H}_{tot}^{\times}(\tau) + \frac{\beta}{2} {\dot {\hat H}}_A^{\circ}(\tau) \right] {\hat K}(\tau).
\label{eq:dW} 
\end{align}
The path (or functional) integral form of  $ {\hat K}(\tau) $ is expressed as
\begin{align}
   K(\xi, \xi',\tau) &= \int \int d\xi_0 d\xi_0' 
\int_{\xi(0)=\xi_0}^{\xi(\tau)=\xi} {\mathcal D}[\xi(t)] \int_{\xi'(0)=\xi'_0}^{\xi'(\tau)=\xi'} {\mathcal D}[\xi'(t)] \nonumber\\
   &\times{\exp}\left[{\frac{i}{\hbar}S_{tot}[\xi; \tau]-\frac{\beta}{2}W[\xi;\tau] } \right] \rho_{tot} (\xi_0, \xi_0', t_0) \nonumber \\
   &\times{\exp}\left[{-\frac{i}{\hbar}S_{tot}[\xi'; \tau]-\frac{\beta}{2}W[\xi';\tau]  }\right],
  \label{eq:Wrho}
\end{align}
where $S_{tot}[\xi; \tau]$ and $W[\xi;\tau] \equiv \int_0^{\tau} dt {\dot H}_A(\xi;t)$ are the total action and the work as a functional of $\xi(t)=(\sigma(t), x_1 (t),x_2 (t),\ldots)$, i.e., the system coordinate appended to the bath coordinate. The above expression implies that $\langle \exp[-\beta W(\tau)] \rangle_{path}$ is obtained as an ensemble average of possible Liouville space pathways $\xi(t)$ and $\xi'(t)$. In the framework of the classical Jarzynski equality,  the paths $\xi(\tau)$ and $\xi'(\tau)$ in the work operators $-\beta W[\xi;\tau]/2 $ and $-\beta W[\xi';\tau]/2 $ are determined as the paths of minimal action for $-iS_{tot}[\xi'; \tau]/\hbar$ and $iS_{tot}[\xi; \tau]/\hbar$, respectively. In the present case, however, the paths $\xi(\tau)$ and $\xi'(\tau)$ are altered by the presence of $-\beta W[\xi;\tau]/2 $ and $-\beta W[\xi';\tau]/2 $ in Eq. (\ref{eq:Wrho}): This violates the condition to satisfy the Jarzynski equality.  Nevertheless, we use Eq. \eqref{eq:P}, because it is natural to assume that the measurement of the work cannot be carried out without disturbing the dynamics of the main system, because it is regarded as small.

For an open quantum system, we can derive the HEOM for Eq. (\ref{eq:Wrho}) using the same procedure as that used to obtain the HEOM for Eq. \eqref{eq:Hamiltonian},\cite{TanimuraJPSJ06,Tanimura89A,TanimuraPRA90,IshizakiJPSJ05,YTJCP14,YTJCP15} because the only difference between the quantum Liouville equation and Eq.\eqref{eq:dW} is the presence of the work operator $ -\beta {\dot {\hat H}_A^{\circ}}(\tau) / 2$ in the latter.
We assume that the spectral density is given by the Drude distribution, $J (\omega) = \eta \gamma^2 \omega / (\omega^2 + \gamma^2)$, where $\eta$ is the SB coupling strength, and $\gamma$ is the inverse correlation time of the bath-induced noise. Then, the noise correlation function takes the form of a linear combination of exponential functions and a delta function: $C(t) = \sum_{k=0}^{L} (c'_k + ic''_k) \gamma_k e^{- \gamma_k t} + 2\Delta_L \delta (t)$, where $c'_k$, $c''_k$, $\gamma_k$, and $\Delta_L$ are constants. 
Then the HEOM for Eq.\eqref{eq:dW} is expressed as
\begin{align}
  \frac{\partial}{\partial t} &{\hat K}_{(n_0,\ldots,n_L)}(t)  \nonumber \\
&= -\left[ \frac{i}{\hbar} {\hat H}_A^{\times}(t)+\frac{\beta}{2}{\dot {\hat H}}_A^{\circ}(t)  +  \Delta_L {\hat \Phi}^2 + \sum_{k=0}^{L} n_k \gamma_k\right] {\hat K}_{(n_0,\ldots,n_L)}(t) \nonumber\\
     &+\sum_{k=0}^{L} n_k {\hat \Theta_k} {\hat K}_{(\ldots,n_k - e_k,\ldots)}(t) + \sum_{k=0}^{L} {\hat \Phi} {\hat K}_{(\ldots,n_k + e_k,\ldots)}(t),
  \label{eq:HEOM-NEW}
\end{align}
where  $\mathbf{e}_k$ is the unit vector along the $k$th direction. Each hierarchical density matrix is specified by the index ${\bf n} = (n_0, \ldots,n_L)$. The density matrix for ${\bf n} = 0$ corresponds to the actual work distribution operator, ${\hat K} (\tau)$.
The initial state is prepared by numerically solving Eq. (\ref{eq:HEOM-NEW}) with the fixed Hamiltonian ${\hat H}_A (t=0) $ until all of the hierarchy elements reach a steady state and then these elements are used as the initial state. 
Because Eq. (\ref{eq:HEOM-NEW}) is identical to the HEOM in the case that $\partial {\hat H}_A (t)/ {\partial t}$ has no explicit time dependence, the steady--state solution of the first hierarchy element is identical to the correlated thermal equilibrium state defined by ${\hat K}_{({\bf n} = 0)}^{eq} = {\rm tr_{B}} \{\exp(-\beta {\hat H} (0))\}/{\rm tr_{A+B}} \{\exp(-\beta {\hat H} (0))\}$.

\begin{figure}[tp]
  \centering
   \includegraphics[width = 8.0cm]{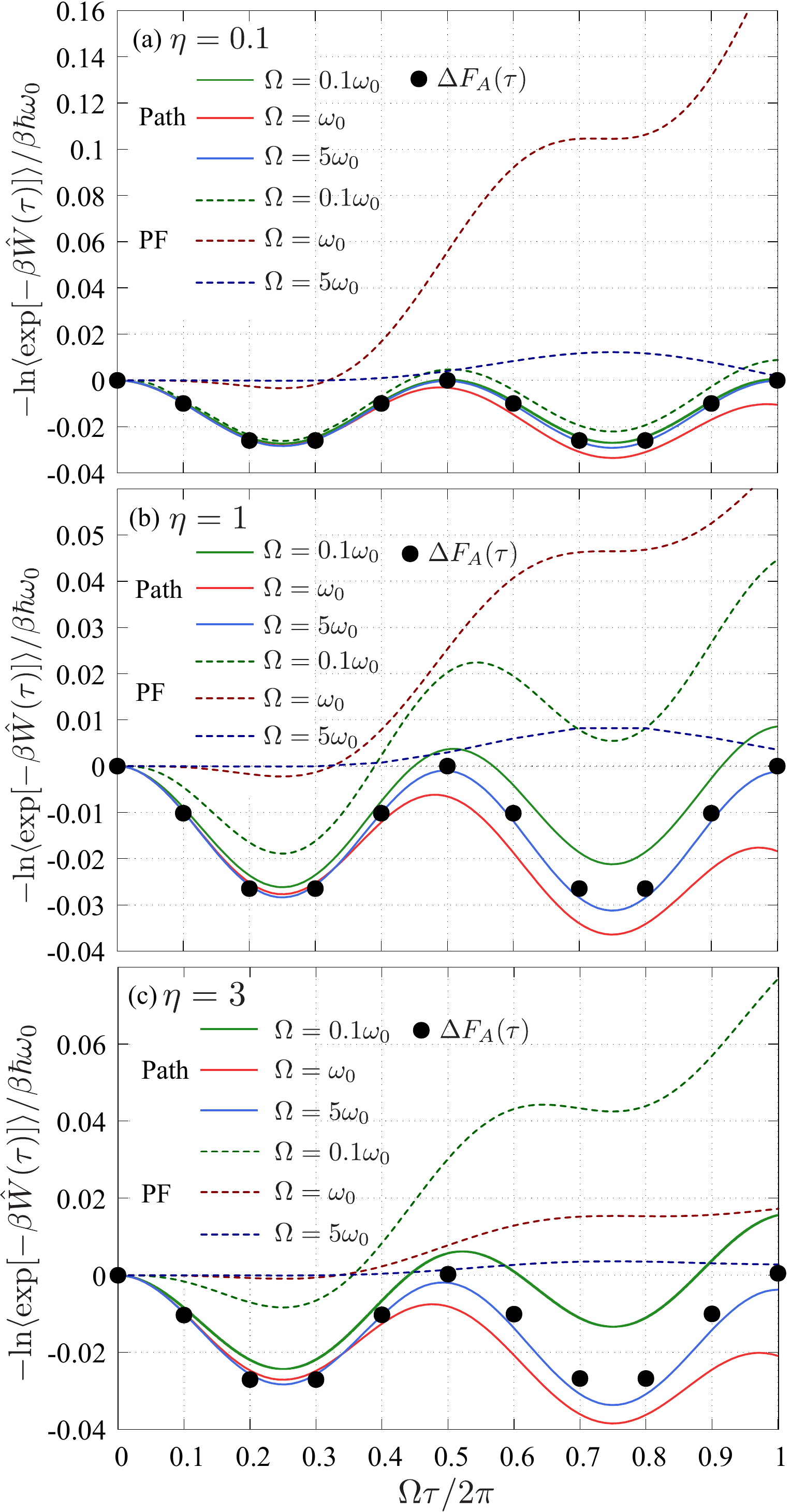}
   \caption{(Color online) The quantity $-{\rm ln}\langle \exp[-\beta W(\tau)] \rangle/\beta$ evalauted as the path-WCF (solid curves) and the PF-WCF (dashed curves), and the change in the free energy, $\Delta F_A(\tau)$ (black dots), under the external perturbation  
${\hat H}_E(t) =(\theta(t) \sin(\Omega t)/4) \hbar \omega_0 (|g\rangle \langle e|+|e\rangle \langle g|)$ are plotted as functions of time for fixed $\beta \hbar \omega_0 = 1$ in (a) the weak ($\eta=0.1$), (b) intermediate ($\eta=1$), and (c) strong ($\eta=3$) SB coupling cases.  The colored dashed and solid curves represent the results for different frequencies: $\Omega=0.1\omega_0$ (red curves), $\Omega=\omega_0$ (green curves) and $\Omega=5\omega_0$ (blue curves). }
   \label{fig:WCF} 
\end{figure}

We now report the results of our numerical computations of the PF-WCF, $\langle \exp[-\beta W(\tau)] \rangle_{PF}$, defined in Eq.\eqref{eq:Hanggi}, and the path-WCF, $\langle \exp[-\beta W(\tau)] \rangle_{path}$, defined in Eq.\eqref{eq:P} (or Eq. \eqref{eq:Wrho}), under the periodic external force described by ${\hat H}_E(t) = \hbar \omega_0 \theta(t) \sin(\Omega t) (|g\rangle \langle e|+|e\rangle \langle g|)/4$, where $\theta(t)$ is the step function and $\Omega$ is the frequency of the external field. The SB interaction is defined as $\hat V =|g\rangle \langle e|+|e\rangle \langle g|$. 
While $\langle \exp[-\beta W(\tau)] \rangle_{path}$ is evaluated using Eq. \eqref{eq:HEOM-NEW},  the procedures for computing $\Delta F_A(\tau)$ and $\langle W(\tau) \rangle$  ($=-\ln\left( \langle \exp[-\beta W(\tau)] \rangle_{PF} \right)/\beta$) on the basis of the HEOM are explained in  Ref. \citeonline{Sakamoto2020JCP}.
The effect of the bath on thermodynamic properties in this SB model is characterized by the SB coupling strength, the bath temperature, and the noise correlation time.\cite{KATO2015,Sakamoto2020JCP,KATO2016}
Throughout this investigation, we fix $\beta \hbar \omega_0 = 1$ and $\gamma=\omega_0$. These values correspond to intermediate temperature and moderately non-Markovian noise. 

	In Fig. \ref{fig:WCF} (a), we display the time dependences of the PF-WCF, path-WCF, 
and $\Delta F_A (\tau)$ for several values of the excitation frequency, $\Omega$, in the weak ($\eta=0.1$) SB coupling case.  
Due to the production of heat, $\Delta Q(\tau)$, the PF-WCF, i.e  $-\ln\left(\langle \exp[-\beta W(\tau)] \rangle_{PF}\right)/\beta=\langle W(\tau) \rangle$,  increases as a function of time in accordance with the first law of thermodynamics,
$\langle W(\tau) \rangle = \Delta F_A(\tau)+ T\Delta S_A(\tau) -\Delta Q(\tau)$. This increase is largest in the resonant case $\Omega = \omega_0$, because the excitation of the system is most efficient there. We calculated $\Delta Q(\tau)$ separately and found that the cycle of the production of heat is similar to that of the WCF.  Because the entropy production $\Sigma_{tot}(\tau) = \Delta S_A(\tau) - \Delta Q(\tau)/T$ exhibits a time lag after the external excitation, we observe a phase delay in the PF-WCF results due to this contribution. The delay is largest at the resonant excitation, $\Omega=\omega_0$, because the entropy production is largest there. The amplitudes of oscillations are suppressed because $-\Delta Q(\tau)$ partially cancels out the contribution from the free energy. For the slowest modulation, $\Omega=0.1\omega_0$, in this weak coupling case, the free energy is almost canceled by the heat production. In this case, the time evolution of the PF-WCF is dominated by $T\Delta S_A(\tau)$.

While the time evolution of the PF-WCF differs significantly from $\Delta F_A (\tau)$, that of the path-WCF is  quite similar. This similarity can be understood as follows. First, note that the ensemble average of $W[\xi;\tau] \equiv \int_0^{\tau} dt {\dot H}_A(\xi;t)$ in Eq. \eqref{eq:Wrho} is taken after the time integration of ${\dot H}_A(\xi;t)$ for a given Liouville path $\xi(t)$. Then, because the contribution of ${\dot H}_A(\xi;t)$ oscillates between positive and negative values rapidly in time, the heat production involved in the definition in Eq. \eqref{eq:Wrho} is suppressed. By contrast, the ensemble average of the PF-WCF is taken step by step in the time integration. Thus, $-\ln\left(\langle \exp[-\beta W(\tau)] \rangle_{PF}\right)/\beta=\langle W(\tau) \rangle$  becomes thermodynamic process function, while $\langle \exp[-\beta W(\tau)] \rangle_{path}$ is not.

In Figs. \ref{fig:WCF} (b) and (c), we present the results for the intermediate and strong SB coupling cases. As pointed out previously,\cite{KATO2015,Saito2008} the efficiency of the heat current is suppressed when the SB coupling becomes strong. Because the effective SB coupling depends on the characteristic time scale of the system, the increase of the PF-WCF is suppressed in the case $\Omega \ge \omega_0$, while it is enhanced in the case $\Omega < \omega_0$.\cite{TanimuraJCP92} In addition, due to the effect of the moderately strong SB coupling, the system closely follows its instantaneous equilibrium state, as the entropy production, $\Sigma_{tot}(\tau)$, is suppressed. As a result, the amplitudes of oscillations and phase delay are small. In Figs. \ref{fig:WCF} (b) and (c), because the eigenenergies of the system are significantly altered by the strong system-bath coupling, the deviation of the time profile of the PF-WCF from $\Delta F_A (\tau)$ increases as $\Omega$ decreases.

For the path-WCF results represented by the solid curves, the deviation becomes larger as the strength of the SB coupling increases. This is because the calculated $\Delta F_A (\tau)$ involves a contribution from the energy of the system part of the SB interaction,\cite{Sakamoto2020JCP} whereas ${\dot H}_A(\xi;t)$ does not include the contribution from $H_I$, which is also time dependent in the reduced description, due to the non-Markovian nature of the noise that arises from the bath. The path-WCF approaches the free energy when the characteristic time scale of the system is shorter than $t_c$, because in this case, the contribution from the SB interaction is insignificant. 

In the weak coupling case considered in Fig. \ref{fig:WCF} (a), the difference between the path-WCF results and the free energy results is large in the resonant excitation case, $\Omega=\omega_0$. This difference has a non-kinetic origin 
arising from the alteration of the Liouville path in Eq. \eqref{eq:Wrho} due to the presence of the work functional $-\beta W[\xi;\tau] \equiv -\beta \int_0^{\tau} dt {\dot H}_A(\xi;t)$. 
 In the case $\Omega =\omega_0$, the contribution of the external perturbation, ${\dot H}_A(\xi;t)$, in the work operator is large, because the cyclic excitation with this excitation strongly perturbs the system dynamics. Thus the paths that should be determined by the total action are altered, and as a result, the calculated path-WCF exhibits time profiles that differ significantly from those in other cases.
For this reason, the path-WCF results also differ significantly from $\Delta F_A (\tau)$ in the low temperature case, because the contribution of $-\beta W[\xi;\tau]$ is larger there (results not shown).

In this paper, we demonstrated a method for extending the Jarzynski equality to the fully quantum regime.  We evaluated the WCF defined in two ways, the PF-WCF and path-WCF, using the numerally rigorous HEOM formalism.  Although the path-WCF agrees with the free energy reasonably well, in particular in a weak SB coupling case or the fast excitation cases, while the PF-WCF exhibits very different time-dependence due to the heat production, the result is not equality but approximation.
This discrepancy arises from the contribution of the SB interaction, which should also play a role in the classical case if the SB coupling strength is comparable to the system energy.  Indeed, if we employ quantum hierarchical Fokker--Planck equations (QHFPEs) for a system described by Wigner distribution functions, we can investigate not only the quantum case but also the classical case by taking the classical limit: We can easily identify purely quantum mechanical effects by comparing the classical and quantum results for the Wigner distribution.\cite{TanimuraJCP92,KatoJPCB13}  

It should be mention that, although here we introduced the path-WCF, this is not physical observable, as seen from Eq. \eqref{eq:dW}.  Moreover, we cannot determined the paths in the functional formalism of Eq. \eqref{eq:Wrho}, due to the limitation introduced by the uncertainty principle.  Thus, in order to evaluate the free energy in the fully quantum regime, the path-WCF is not practical. Instead, Eq. \eqref{eq:Fchange} should be used to evaluate the free energy.

Although the present investigation is limited to spin-Boson systems for the specific definitions of the WCF, the applicability of our approach based on the HEOM formalism is in fact more general. Indeed, the same approach can be applied to all of the systems to which the HEOM formalism has been previously applied.\cite{TanimuraJPSJ06,YTperspective} Different definitions of the WCF should also be examined. We leave such extensions to future studies to be carried out in the context of the fluctuation theorem.


\begin{thebibliography}{99}
\bibitem{JarzynskiPRL97} C. Jarzynski, Phys. Rev. Lett. \textbf{78}, 2690 (1997).
\bibitem{Jarzynski04} C. Jarzynski, J. Stat. Mech. (2004)  P09005.
\bibitem{JarzynskiAnnu11} C. Jarzynski, {Annu. Rev. Condens. Matter Phys.} \textbf{2} (2011) 321. 
\bibitem{CrooksPRE99} G. E. Crooks, Phys. Rev. E \textbf{60}, 2721 (1999).
\bibitem{SeifertPRL05} U. Seifert, Phys. Rev. Lett. \textbf{95}, 040602 (2005).
\bibitem{MaiPRE07} T. Mai, A. Dhar, Phys. Rev. E \textbf{75}, 061101 (2007).
\bibitem{SeifertJSM07} T. Speck, U. Seifert, J. Stat. Mech. L09002 (2007).
\bibitem{SaitoPRB08} K. Saito, Y. Utsumi, Phys. Rev. B \textbf{78}, 115429 (2008).
\bibitem{EvansPRL08} S. R. Williams, D. J. Searles, D. J. Evans, Phys. Rev. Lett. \textbf{100}, 250601 (2008).
\bibitem{Liphardt02} J. Liphardt, S. Dumont, S. B. Smith,I. Tinoco, Jr., C. Bustamante, Science \textbf{296}, 1832 (2002).
\bibitem{DouarcheEuro05} F. Douarche, S. Ciliberto, A. Petrosyan, I. Rabbiosi Europhys. Lett. \textbf{70}, 593 (2005).
\bibitem{CollinNature05} D. Collin, F. Ritort, C. Jarzynski, S. B. Smith, I. Tinoco, Jr., C. Bustamante, Nature \textbf{437}, 231, (2005).
\bibitem{BlickePRL06} V. Blickle, T. Speck, L. Helden, U. Seifert, C. Bechinger  Phys. Rev. Lett. \textbf{96}, 070603 (2006).
\bibitem{RitortJCP09} A. Mossa, S. de Lorenzo, J. M. Huguet, F. Ritort, J. Chem. Phys. \textbf{130}, 234116 (2009).
%
\bibitem{Alonso} I. de Vega and D. Alonso, {Rev. Mod. Phys.} \textbf{89} (2017) 015001.
\bibitem{Kurchan} J. Kurchan, e-print {arXiv:cond-mat/0007360} (2000).
\bibitem{Tasaki} H. Tasaki, e-print {arXiv:cond-mat/0009244} (2000).
\bibitem{YukawaJPSJ00} S. Yukawa, {J. Phys. Soc. Jpn.} \textbf{69} (2000)  2367.
\bibitem{CrooksStat08} G. E. Crooks, {J. Stat. Phys.}   (2008) P10023.
\bibitem{Campisi09} M. Campisi, P. Talkner and P. H\"{a}nngi, {Phys. Rev. Lett.} \textbf{102} (2009)  210401.
\bibitem{Campisi11} M. Campisi, P. H{\"a}nggi, and P. Talkner, {Rev. Mod. Phys.} \textbf{83} (2011) 771.
\bibitem{Mukamel} M. Esposito, U. Harbola, and S. Mukamel, {Rev. Mod. Phys.} \textbf{81}  (2019) 1665.
\bibitem{Hanggi2020}P. Talkner and P. H{\"a}nggi {Rev. Mod. Phys.} \textbf{92} (2020) 041002.
\bibitem{KATO2016}A. Kato and Y. Tanimura,  {J. Chem. Phys.} {\bf 145}  (2016) 224105.
\bibitem{Sakamoto2020JCP}S. Sakamoto and Y. Tanimura, J. Chem. Phys. \textbf{153} (2020) 234107.
\bibitem{YTperspective} Y. Tanimura, J. Chem. Phys.  \textbf{153} (2020) 020901.
 \bibitem{TanimuraJPSJ06}Y. Tanimura, J. Phys. Soc. Jpn. {\bf75} (2006)  082001.
\bibitem{Tanimura89A} Y. Tanimura and R. Kubo, J. Phys. Soc. Jpn. {\bf 58} (1989) 101.
 \bibitem{TanimuraPRA90} Y. Tanimura, Phys. Rev. {\bf A41} (1990) 6676. 
 \bibitem{IshizakiJPSJ05} A. Ishizaki and Y. Tanimura, J. Phys. Soc. Jpn. \textbf{74} (2005) 3131.
 \bibitem{YTJCP14} Y. Tanimura, J. Chem. Phys. {\bf 141} (2014) 044114.
 \bibitem{YTJCP15} Y. Tanimura, J. Chem. Phys. {\bf 142} (2015) 144110. 
 \bibitem{Talkner16} B. P. Venkatesh, G. Watanabe, P. Talkner, New J. Phys. {\bf 17} (2017) 075018. 
 \bibitem{HanggiW} P. Talkner, E. Lutz, P. H\"{a}nggi, Phys. Rev. E 75 (2007)  050102(R).
\bibitem{KATO2015}A. Kato and Y. Tanimura,  {J. Chem. Phys,} {\bf 143} (2015) 064107.
\bibitem{Saito2008} K. Saito, {Europhys. Lett.} \textbf{83} (2008) 50006.
\bibitem{TanimuraJCP92}Y. Tanimura and P. G. Wolynes, {J. Chem. Phys.} {\bf 96} (1992)  8485.
\bibitem{KatoJPCB13}A. Kato and Y. Tanimura, {J. Phys. Chem. B} {\bf 117} (2013) 13132. 
\end{thebibliography}
\end{document}